\documentclass[conference,a4paper]{IEEEtran}
\IEEEoverridecommandlockouts
\usepackage{cite}
\usepackage{amsmath,amssymb,amsfonts,bm}
\usepackage{graphicx}
\usepackage{textcomp}
\usepackage{xcolor}
\usepackage[nolist]{acronym}
\usepackage{tikz,pgfplots}
\usepackage[font=footnotesize]{caption}
\usepackage{algorithm}
\usepackage{algpseudocode}
\usepackage{siunitx}
\usepackage{tabularx, booktabs} 
\usepackage[T1]{fontenc}
\usepackage{multirow}
\usepackage[inline]{enumitem}
\usepackage{soul}
\usepackage[hidelinks]{hyperref}

\DeclareSIUnit{\belmilliwatt}{Bm}
\DeclareSIUnit{\dBm}{\deci\belmilliwatt}

\def\BibTeX{{\rm B\kern-.05em{\sc i\kern-.025em b}\kern-.08em
    T\kern-.1667em\lower.7ex\hbox{E}\kern-.125emX}}

\begin{acronym}
\small
\acro{5G}{fifth-generation}
\acro{6G}{sixth-generation}
\acro{ACK}{Acknowledgment}
\acro{AP}{access point}
\acro{AUD}{active users detection}
\acro{AUE}{active users enumeration}
\acro{AUI}{active users identification}
\acro{AUEI}{active users enumeration and identification}
\acro{AMP}{approximate message passing}
\acro{ANN}{artificial neural network}
\acro{AWGN}{additive white Gaussian noise}
\acro{ACRDA}{asynchronous contention resolution diversity ALOHA}
\acro{AER}{activity error rate}
\acro{AoI}{age of information}
\acro{AMP}{approximate message passing}
\acro{AWGN}{additive white Gaussian noise}
\acro{BAC}{binary adder channel}
\acro{BCH}{Bose–Chaudhuri–Hocquenghem}
\acro{BEC}{binary erasure channel}
\acro{BN}{burst node}
\acro{BPR}{Bar-David, Plotnik, Rom}
\acro{BPSK}{binary phase shift keying}
\acro{BS}{base station}
\acro{BIHT}{block iterative hard thresholding}
\acro{CDF}{cumulative distribution function}
\acro{CDMA}{code division multiple access}
\acro{CF-mMIMO}{cell-free massive MIMO}
\acro{CN}{check node}
\acro{CNN}{convolutional neural network}
\acro{CPU}{central processing unit}
\acro{CRA}{coded random access}
\acro{CRC}{cyclic redundancy check}
\acro{CRDSA}{contention resolution diversity slotted ALOHA}
\acro{CS}{compressed sensing}
\acro{CSA}{coded slotted ALOHA}
\acro{CSI}{channel state information}
\acro{DE}{density evolution}
\acro{DFT}{Discrete Fourier Transform}
\acro{DL}{deep learning}
\acro{DMC}{discrete memoryless channel}
\acro{DNN}{deep neural network}
\acro{DRL}{deep reinforcement learning}
\acro{DS}{delay spread}
\acro{ECRA}{enhanced contention resolution ALOHA}
\acro{EGF}{exponential generating function}
\acro{eLU}{exponential linear unit}
\acro{eMBB}{enhanced mobile broad-band}
\acro{EXIT}{extrinsic information transfer}
\acro{FER}{frame error rate}
\acro{FLOP}{floating point operation}
\acro{G-CSA}{graph-defined CSA}
\acro{G-IRSA}{graph-defined IRSA}
\acro{GMAC}{Gaussian multiple access channel}
\acro{gNB}{Next Generation NodeB}
\acro{GF}{grant-free}
\acro{GLRT}{generalized likelihood ratio test}
\acro{GP}{geometric programming}
\acro{HLL}{high-low-low}
\acro{HTC}{human-type communication}
\acro{IC}{interference cancellation}
\acro{IDFT}{inverse discrete Fourier transform}
\acro{IFSC}{intra-frame spatial coupling}
\acro{IHT}{iterative hard thresholding}
\acro{i.i.d.}{independent and identically distributed}
\acro{IoT}{Internet of Things}
\acro{IRA}{irregular repetition ALOHA}
\acro{IRSA}{irregular repetition slotted ALOHA}
\acro{JADCE}{joint activity detection and channel estimation}
\acro{JPCPA}{joint power control and pilot assignment}
\acro{KPI}{key performance indicator}
\acro{LDPC}{low-density parity-check}
\acro{LDS}{low-density signature}
\acro{LOS}{line-of-sight}
\acro{LRT}{likelihood ratio test}
\acro{LSTM}{long short-term memory}
\acro{LT}{Luby transform}
\acro{LTE}{Long Term Evolution}
\acro{mIoT}{massive Internet of Things}
\acro{MAC}{medium access control}
\acro{ML}{maximum likelihood}
\acro{MAP}{maximum a posteriori}
\acro{MDS}{maximum distance separable}
\acro{MGF}{moment generating function}
\acro{MIMO}{multiple-input multiple-output}
\acro{MMSE}{minimum mean square error}
\acro{MMA}{massive multiple access}
\acro{MMV}{multiple measurement vector}
\acro{MTC}{machine-type communication}
\acro{mMTC}{massive machine-type communication}
\acro{mMIMO}{massive multiple-input multiple-output}
\acro{MPR}{multi-packet reception}
\acro{M-QAM}{$\mathsf{M}$-Quadrature Amplitude Modulation}
\acro{QAM}{Quadrature Amplitude Modulation}
\acro{NOMA}{non-orthogonal multiple access}
\acro{OMA}{orthogonal multiple access}
\acro{OFDM}{Orthogonal Frequency-Division Multiplexing}
\acro{PEG}{progressive edge growth}
\acro{MPR}{multi-packet reception}
\acro{mMTC}{massive machine-type communications}
\acro{MLS}{maximum length sequence }
\acro{MRC}{maximal ratio combining}
\acro{MUD}{multi-user detection}
\acro{NMSE}{nomalized mean square error}
\acro{NLOS}{non-line-of-sight}
\acro{NOMA}{non-orthogonal multiple access}
\acro{OMP}{orthogonal matching pursuit}
\acro{PAB}{payload aided-based}
\acro{PAPR}{peak-to-average power ratio}
\acro{PB}{power balance}
\acro{PDF}{probability density function}
\acro{PDMA}{power domain multiple access}
\acro{PD-NOMA}{power-domain NOMA}
\acro{PGF}{probability generating function}
\acro{PMF}{probability mass function}
\acro{PHY}{physical}
\acro{PL}{path loss}
\acro{PLR}{packet loss rate}
\acro{PN}{pseudo-noise}
\acro{PRACH}{physical random access channel}
\acro{PRCE}{perfect replica channel estimation}
\acro{PU}{power unbalance}
\acro{QAM}{Quadrature Amplitude Modulation}
\acro{QPSK}{quadrature phase-shift keying}
\acro{RA}{random access}
\acro{RF}{radio-frequency}
\acro{ReLU}{rectified linear unit}
\acro{RL}{reinforcement learning}
\acro{RMSE}{root-mean-square error}
\acro{RNN}{recurrent neural network}
\acro{ROC}{receiver operating characteristics}
\acro{RS}{Reed Solomon}
\acro{RSC}{randomized spatial coupling}
\acro{r.v.}{random variable}
\acro{SA}{slotted ALOHA}
\acro{SC}{spatial coupling}
\acro{SCMA}{sparse code multiple access}
\acro{SE}{spectral efficiency}
\acro{SIC}{successive interference cancellation}
\acro{SINR}{signal-to-interference-plus-noise ratio}
\acro{SIS}{successive interference subtraction}
\acro{SISO}{single input, single output}
\acro{SN}{sum node}
\acro{SNB}{squared-norm based}
\acro{SNR}{signal-to-noise ratio}
\acro{SPCPA}{Separate Power Control and Pilot Assignment}
\acro{SSC}{spaced spatial coupling}
\acro{SVD}{singular value decomposition}
\acro{ToA}{time of arrival}
\acro{TDD}{time-division duplex}
\acro{TDL}{tapped delay line}
\acro{URLLC}{ultra-reliable and low-latency communications}
\acro{VF}{virtual frame}
\acro{ZC}{Zadoff-Chu}
\end{acronym}
\newcommand{\mat}[1]{\bm{#1}} 
\DeclareMathOperator*{\diag}{diag}

\definecolor{cobaltblue}{rgb}{0,0.27,0.67}
\definecolor{codegreen}{rgb}{0,0.6,0.3}
\definecolor{codegray}{rgb}{0.5,0.5,0.5}
\definecolor{codepurple}{rgb}{0.58,0,0.82}
\definecolor{codered}{rgb}{0.7,0,0}
\definecolor{codeblue}{rgb}{0,0,0.7}
\definecolor{backcolour}{rgb}{0.95,0.95,0.98}
\definecolor{blue(ryb)}{rgb}{0.01, 0.28, 1.0} 
\definecolor{carrotorange}{rgb}{0.89, 0.53, 0.09}

\makeatletter
\newcommand*{\rom}[1]{\expandafter\@slowromancap\romannumeral #1@}
\makeatother
\newcolumntype{Y}{>{\centering\arraybackslash}X}

\begin{document}

\title{Blind User Activity Detection for Grant-Free Random Access in Cell-Free mMIMO Networks\\
}

\author{\IEEEauthorblockN{Muhammad Usman Khan, Enrico Testi, Marco Chiani, Enrico Paolini}
\IEEEauthorblockA{CNIT/WiLab, DEI, University of Bologna, Italy\\
Email: \{muhammadusman.khan8, enrico.testi, marco.chiani, e.paolini\}@unibo.it}}

\maketitle

\begin{abstract}
\Ac{CF-mMIMO} networks have recently emerged as a promising solution to tackle the challenges arising from next-generation \acl{mMTC}.
In this paper, a fully grant-free \ac{DL}-based method for user activity detection in \ac{CF-mMIMO} networks is proposed.
Initially, the known non-orthogonal pilot sequences are used to estimate the channel coefficients between each user and the access points. Then, a deep \acl{CNN} is used to estimate the activity status of the users. The proposed method is ``blind'', i.e., it is fully data-driven and does not require prior large-scale fading coefficients estimation. Numerical results show how the proposed \ac{DL}-based algorithm is able to merge the information gathered by the distributed antennas to estimate the user activity status, yet outperforming a state-of-the-art covariance-based method.
\end{abstract}

\begin{IEEEkeywords}
Cell-free mMIMO, deep learning, user activity detection, grant-free access, mMTC communications
\end{IEEEkeywords}

\acresetall

\section{Introduction}
The ever-growing integration of \ac{IoT} technologies across diverse sectors has led to a surge in the number of interconnected smart devices, catalyzing the emergence of \ac{mMTC} systems \cite{Bockelmann2016:Massive}.
In the forthcoming 6G era, \ac{mMTC} systems will face more stringent demands in scalability and device battery lifetime compared to 5G, necessitating reliability and latency levels tailored for specific use cases \cite{Pokhrel2020:Towards}.
Consequently, to tackle the challenges arising from increasing network densities and heterogeneous service requirements for next-generation \ac{mMTC}, the development of advanced MAC and PHY layer protocols, as well as the exploration of novel network architectures is crucial \cite{KhaTesPao:24}.
Among innovative architectures, \ac{CF-mMIMO} networks have recently emerged as a promising alternative to traditional networks \cite{Ngo2017:Cell}. This architectural paradigm, envisioned as a viable solution for future 6G networks, diverges from a cell-based structure by distributing a massive number of \acp{AP} across the network area. These \acp{AP} are coordinated by one or more powerful \acp{CPU}, shifting the system from a base station-centric to a user-centric model. This enables users to access the network via nearby \acp{AP}, thereby maximizing the quality of service.

The sporadic nature of \ac{mMTC} communications, characterized by devices intermittently waking up to transmit short payloads, underscores the inefficiency of conventional grant-based protocols due to their substantial signaling overhead. This inefficiency not only limits the overall network performance, but also leads to unnecessary energy consumption on the devices' side. Recently, various grant-free random access protocols have been proposed to overcome such limitations \cite{RioPanMiu:J21,TesTraPao:J24}. In such schemes, each device is usually assigned a unique pilot sequence, which is used to perform channel estimation and identify which users are active.
To support a large number of \ac{mMTC} devices and due to the limited channel coherence interval length, assigning orthogonal pilot sequences to the devices is not feasible. For this reason, unique non-orthogonal pilot sequences are usually assigned to each user, inducing severe co-channel interference and making channel estimation and user activity detection challenging problems.

In the last decade, several works addressed the user activity detection problem in single and multi-cell scenarios \cite{MonWolBoc:C15,DuDonChe:J17,SenLar:J18, HagJunCai:18, BaiAiChe:C19, ZhaLiHua:J19,KhaPaoChi:22}.
In \cite{MonWolBoc:C15,DuDonChe:J17}, the activity detection problem is formulated as a \ac{CS} problem and solved using greedy pursuit algorithms. Random and structured sparsity learning-based multi-user detection was studied in \cite{SenLar:J18}. The covariance-based approach detailed in \cite{HagJunCai:18} estimates user activity in a cell-based scenario. \Ac{DL}-based user activity detection approaches are proposed in \cite{BaiAiChe:C19,ZhaLiHua:J19,KhaPaoChi:22}.

To the best of our knowledge, only a few recent works address the problem of active user detection in cell-free networks \cite{GanBjoLar:J21, BaiLar:22}. In \cite{GanBjoLar:J21}, the maximum likelihood user activity detection problem for grant-free random access in cell-free networks is formalized. The authors propose a clustering-based activity detection algorithm that leverages the distributed nature of cell-free systems and the computational capability of the \ac{CPU}.
In \cite{BaiLar:22}, the authors utilize a distributed \ac{AMP} algorithm for joint activity detection and channel estimation. Both approaches necessitate the estimation of large-scale fading coefficients between the \acp{AP} and the devices before active user detection. Estimating large-scale fading coefficients in grant-free networks with cell-free architectures poses substantial challenges and may result in imperfect estimates \cite{Fengler2021}. Additionally, non-perfect large-scale fading coefficient estimation can compromise active user detection performance.

In this paper, a fully grant-free \ac{DL}-based method for user activity detection in \ac{CF-mMIMO} networks that does not necessitate prior large-scale fading coefficients and noise power knowledge is proposed. The aggregate received symbols at the \ac{CPU} and the known non-orthogonal pilot sequences are first used to estimate the channel coefficients between each user and each antenna of the \acp{AP}. Then, a deep \ac{CNN} is used to estimate the activity status of the users.
The proposed \ac{DL} algorithm is able to merge the information gathered by the distributed antennas and consistently detect which users are active.

The rest of the paper is organized as follows. Section~\ref{sec:system_model} outlines the system model. The \ac{DL}-based approach for user activity detection in \ac{CF-mMIMO} is described in Section~\ref{sec:DL}. Simulation setup along with the numerical results are provided in Section~\ref{sec:results}. Conclusions are drawn in Section~\ref{sec:conclusion}.

Matrices, vectors, and scalars are represented by boldface uppercase, boldface lowercase, and lowercase letters, respectively. The fields of real and complex numbers are denoted by $\mathbb{R}$ and $\mathbb{C}$, respectively. The operations $(\cdot)^T$ and $(\cdot)^H$ denote the transpose, and conjugate transpose, respectively. The Euclidean norm operator is defined as $\|\cdot\|^2$, respectively. The normal and circularly-symmetric complex normal distributions with mean $0$ and variance $\sigma^2$ are denoted by $\mathcal{N}(0,\sigma^2)$ and  $\mathcal{CN}(0,\sigma^2)$, respectively. 
The convolution operation between matrices $\bm{A}$ and $\bm{B}$ is indicated as $\bm{A}*\bm{B}$. The $J \times J$ identity matrix is denoted by $\mat{I}_J$.

\section{System Model}\label{sec:system_model}
\begin{figure}[t]
    \centering
    \includegraphics[width=.8\columnwidth]{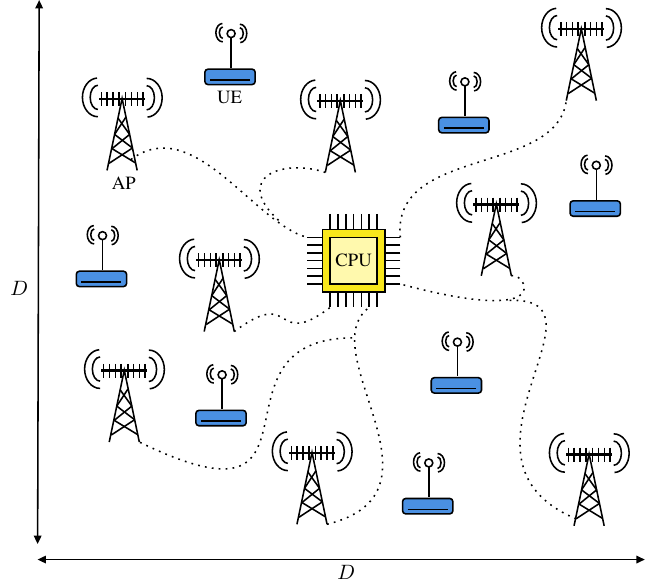}
    \caption{An illustration of a cell-free massive MIMO network. }
    \label{fig:scenario}
\end{figure}
\begin{figure*}[tb]
    \centering
    \input{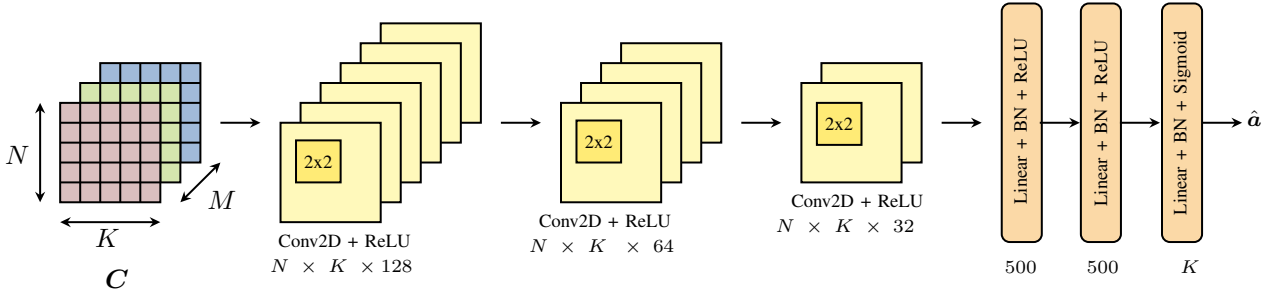}
    \caption{CNN architecture for user activity detection in a \ac{CF-mMIMO} scenario. 
    The number of filters and neurons is specified beneath each layer. For example, the first convolutional layer employs $128$ filters, and the first linear layer consists of $500$ neurons. }
    \label{fig:arch}
\end{figure*}
We consider a \ac{CF-mMIMO} network with $M$ \acp{AP} each equipped with $N$ antennas serving $K$ single-antenna users randomly distributed in an area of $D \times D$ m\textsuperscript{2}, as illustrated in Fig.~\ref{fig:scenario}. The \acp{AP} are connected to a \ac{CPU} through fronthaul links. We assume that the users are synchronized and are served by all \acp{AP} in the same time-frequency resources. Each user activates with probability $\epsilon \ll 1$. The vector of received symbols $\mat{y}_{mn} \in \mathbb{C}^{\tau \times 1}$ at the $n$th antenna of the $m$th \ac{AP} is
\begin{align}
    \mat{y}_{mn} &= \sqrt{\rho} \sum_{k =1}^K  a_k  g_{mnk} \mat{\phi}_k + \mat{w}_{mn} 
\end{align}
where 
\begin{itemize}
    \item $\rho$ is the nodes' transmit power; 
    \item $a_k  \in \{0,1\}$ is the activity indicator of the $k$th device, with $a_k = 0/1$ for inactive/active devices, respectively;
    \item $g_{mnk}= \sqrt{\beta_{mk}} h_{mnk}$ is the channel gain between the $k$th user and the $n$th antenna of the $m$th \ac{AP}, where $\beta_{mk}$ is the large-scale fading incorporating both path-loss and log-normal shadowing, and $ h_{mnk} \sim \mathcal{CN}(0,1) $ is the small-scale fading. The large-scale fading coefficient is $\beta_{mk} = 10^{\sigma_\mathrm{sh} s_{mk}/{10}}/\mathrm{PL}_{mk}$, where $\mathrm{PL}_{mk}$ is the path-loss from the $k$th user to the $m$th \ac{AP}, $\sigma_\mathrm{sh}$ is the shadowing intensity, and $s_{mk} \sim \mathcal{N}(0,1)$;
    \item $\mat{\phi}_k \in \mathbb{C}^{\tau \times 1}$ is the pilot sequence associated with the $k$th user, distributed as $\mat{\phi}_k \sim \mathcal{CN}(0, \mat{I}_{\tau})$;   
    \item $\mat{w}_{mn} \in \mathbb{C}^{\tau \times 1}$ is the vector of \ac{i.i.d.} noise samples, distributed as $\mathcal{CN} (0,\mat{I}_{\tau})$.
\end{itemize}
The received symbols at the $m$th \ac{AP} can be expressed as 
\begin{align}
    \mat{Y}_m &= \sqrt{\rho} \mat{\Phi}  \mat{A} \mat{G}_m + \mat{W}_m 
\end{align}
where $\mat{\Phi} = [\mat{\phi}_1, \dots  ,\mat{\phi}_K] \in \mathbb{C}^{\tau \times K}$ is the aggregate of the pilot symbols transmitted by the users, $\mat{A} = \diag(a_1 , \ldots, a_K) \in \mathbb{R}^{K \times K}$ is a diagonal matrix that contains the user activity indicators, and $\mat{W}_m = [\mat{w}_{m1}, \ldots ,\mat{w}_{mN}] \in \mathbb{C}^{\tau \times N}$ is the matrix of the aggregate noise samples at \ac{AP} $m$. 
The channel gains between the $K$ users and the $m$th \ac{AP} are stacked in matrix 
$\mat{G}_m = [\mat{g}_{m1},\ldots, \mat{g}_{mN}]$ with $\mat{g}_{mn}=[g_{mn1},\ldots,g_{mnK}]^T$.

The received symbols at the \acp{AP} are forwarded to the \ac{CPU} for activity detection. At the \ac{CPU}, the symbols are aggregated as
\begin{align}
    \mat{Y} &= 
    \begin{bmatrix}
        \mat{Y}_1 \\
        \mat{Y}_2 \\
        \vdots \\
        \mat{Y}_M
    \end{bmatrix}
    = \begin{bmatrix}
           \sqrt{\rho}\mat{\Phi}\mat{A}\mat{G}_1 \\
           \sqrt{\rho} \mat{\Phi}\mat{A}\mat{G}_2 \\
           \vdots \\
           \sqrt{\rho} \mat{\Phi}\mat{A}\mat{G}_M \\
        \end{bmatrix}
        + \begin{bmatrix}
           \mat{W}_1\\
           \mat{W}_2 \\
           \vdots \\
           \mat{W}_M \\
        \end{bmatrix} \, .
\end{align}
Let us denote by $\mat{Q} \in \mathbb{R}^{\tau M \times \tau M}$ the matrix
\begin{align}
    \mat{Q}\!&=\! 
     \renewcommand{\arraystretch}{0.9} 
    \setlength{\arraycolsep}{2pt} 
    \begin{bmatrix}
        \rho\mat{\Phi}\mat{A}\mat{B}_1\mat{\Phi}^H & \mat{0} & \hdots & \mat{0} \\
        \mat{0} &  \rho\mat{\Phi}\mat{A}\mat{B}_2\mat{\Phi}^H & \hdots & \mat{0} \\
        \vdots & \vdots & \ddots & \vdots \\
        \mat{0} & \mat{0} & \hdots &  \rho\mat{\Phi}\mat{A}\mat{B}_M\mat{\Phi}^H
    \end{bmatrix}\!+\!\sigma_n^2 \mat{I}_{\tau M}
\end{align}
where $\mat{B}_m = \diag(\beta_{m1}, \ldots, \beta_{mK} )$. Due to the independent small-scale fading affecting the antennas of all the \acp{AP}, the columns of $\mat{Y}$ are also independent, with each column distributed as $\mat{Y}(:,i) \sim \mathcal{CN}(\mat{0},\mat{Q}), \, \forall i= 1,\ldots, N$. 
Let us define vector $\mat{\eta} = \rho \, [a_1, \ldots, a_K]$ and $\mat{Q}_m = \rho \mat{\Phi} \mat{A} \mat{B}_m \mat{\Phi}^H + \sigma_n^2 \mat{I}_{\tau}$.
Leveraging the block-diagonal structure of $\mat{Q}$, the likelihood of $\mat{Y}$ given $\mat{\eta}$ can be expressed as \cite{GanBjoLar:J21}
\begin{align}\label{eq:ML}
    p(\mat{Y} | \mat{\eta}) &= \prod_{m=1}^M \prod_{n=1}^N \frac{1}{|\pi \mat{Q}_m|} \exp\left(-\mat{y}_{mn}^H \mat{Q}_m^{-1}\mat{y}_{mn}\right) \nonumber \\
    &=  \prod_{m=1}^M  \frac{1}{|\pi \mat{Q}_m|^N} \exp\left(-\mathrm{Tr} ( \mat{Q}_m^{-1} \mat{Y}_{m} \mat{Y}_{m}^H  ) \right).
\end{align}
Assuming that the large-scale fading coefficients between all the \acp{AP} and the users, i.e., matrices $\mat{B}_1$, $\mat{B}_2$, $\ldots$, $\mat{B}_M$, are perfectly known at the \ac{CPU}, the user activity detection problem can be solved by finding the vector $\mat{\eta}$ that maximizes \eqref{eq:ML}. However, accurate estimation of the large-scale fading coefficients in grant-free networks operating with cell-free architectures may be unfeasible. For this reason, in the following we propose a data-driven method for user activity detection that does not rely on knowledge of the large-scale fading coefficients.

\section{Deep Learning-based Approach}\label{sec:DL}

In this section, the proposed \ac{DL}-based algorithm for user activity detection, composed by a pre-processing stage and a \ac{CNN}, is presented.

\subsection{Pre-processing}
In the pre-processing stage, we compute the channel estimates at each antenna of each \ac{AP} by projecting the corresponding received symbols onto the known normalized pilot sequences, denoted as $\tilde{\mat{\Phi}} = \mat{\Phi}/\| \mat{\Phi} \|^2$. The channel estimates at the $m$th \ac{AP}, denoted by $\hat{\mat{G}}_m \in \mathbb{R}^{K \times N}$, are obtained as
\begin{align} \label{eq:input}
    \hat{\mat{G}}_m = \frac{ \tilde{\mat{\Phi}}^H \mat{Y}_m  }{\sqrt{\rho}}.
\end{align}
Then, the channel estimates at each \ac{AP} are organized in tensor $\mat{C} \in \mathbb{R}^{N \times K \times M}$, which is fed as input to the \ac{CNN}.

\subsection{CNN Architecture}
In the following, we present the \ac{CNN} designed for user activity detection in a \ac{CF-mMIMO} scenario. To identify the best network architecture for our purpose, we investigated multiple layouts by varying various parameters, such as the number of convolutional/linear layers, filter/kernel sizes, and activation functions. After careful consideration, based on the performance and computational complexity trade-off, we selected the architecture illustrated in Fig.~\ref{fig:arch}. The first layer of the network is a 2D-convolutional layer, yielding as output matrix $\mat{S}$. The entries of $\mat{S}$, $s_{ij}$, are the results of convolution operations, and expressed mathematically as
\begin{align}\label{eq:conv1}
    s_{ij} &= \mat{C}*\mat{F} = \sum_{n=1}^N \sum_{k=1}^K c_{nk}  f_{i-n, j-k} \nonumber \\ 
    &  i = 1,\ldots , \bigg \lfloor \frac{N-F+2P}{T} +1   \bigg \rfloor \nonumber \\
    &  j = 1,\ldots , \bigg \lfloor\frac{K-F+2P}{T} +1 \bigg  \rfloor
\end{align}
where $\mat{F} \in \mathbb{R}^{F \times F}$ is the kernel/filter, while $P$ and $T$ denote the padding and stride sizes, respectively.  
In the proposed architecture, we employ three 2D-Convolution layers consisting of $128$, $64$, and $32$ filters, respectively. We use same padding for the convolutional layers, such that the convolution input and output sizes are equal. We employ a kernel/filter of size $2 \times 2$ for $N > 1$ and a filter of size $1 \times 2$ for $N=1$, with a stride of $1$.

After the 2D-convolution layers, we employ a set of linear layers consisting of multiple neurons, whose output is defined as
    \begin{align}\label{eq:linear}
    \mat{z} &= \mat{W} \mat{\alpha} + \mat{\delta} 
    \end{align}
where  $\mat{W} \in \mathbb{R}^{\mathrm{out} \times \mathrm{in}}$, $\mat{\alpha} \in \mathbb{R}^{\mathrm{in} \times 1}$ and $\mat{\delta} \in \mathbb{R}^{\mathrm{out} \times 1}$ represent the weight matrix, input, and bias, respectively \cite{GooBenCou:16}. Here, $\mathrm{out}$ and $\mathrm{in}$ represent the number of layer's output and input features, respectively. The number of neurons in each layer is specified in Fig.~\ref{fig:arch}. Specifically, the first and second linear layers consist of $500$ neurons each, whereas the last linear layer size corresponds to the number of users. Each output neuron is responsible for detecting the activity status of a user, i.e.,  determining whether it is active or inactive.

To train the \ac{CNN}, we split the training dataset into $N_\text{B}$ mini-batches, each containing $B$ samples. Let us denote by $z_i$ the output of the linear layer in \eqref{eq:linear}, computed from the $i$th element of a training example of one mini-batch. 
The batch normalization layer transforms the input making it zero mean and unit variance, and then scales it using the trainable parameters $\gamma$ and $\xi$. The output of such a layer is defined as
\begin{align}\label{eq:BN}
   r_i = \mathrm{BN}(z_i) = \frac{\gamma (z_i  - \mu_{i}) }{\sqrt{\sigma_i^2}} + \xi
\end{align}
where $\mu_i$ and $\sigma_i^2$ represent the moving mean and variance computed over $i$th element across all the training examples in the mini-batch, respectively \cite{Cho:15}.
The activation function plays a critical role in introducing non-linearity into the network and is applied element-wise to the input. We employ \ac{ReLU} activation function \cite{GooBenCou:16}, defined as
\begin{align}
    \mathrm{ReLU}(r_i) = \max(0, r_i) \, .
\end{align}
The activity detection is handled as a multi-label classification  problem for which we leverage the sigmoid function
\begin{align}\label{eq:sigmoid}
    \hat{a}_i = \mathrm{Sigmoid}(r_i) =  \frac{1}{1+e^{-r_i}}, \qquad i=1, \ldots, K
\end{align}
where $\hat{a}_i$ represents the likelihood of user $i$ being active. 

\subsection{DNN Training}
The proposed \ac{CNN} is trained to solve a multi-label classification problem employing binary cross-entropy loss, defined as
\begin{align}\label{eq:bloss}
    \mathcal{J} ( \mat{a}, \mat{\hat{a}} ) = - \sum_{i =1}^{K} a_i \log \hat{a}_i + (1 -  a_i ) \log  (1 - \hat{a}_i)
\end{align}
where $\mat{a}=[a_1,a_2,\ldots,a_K]$ and $\hat{\mat{a}}=[\hat{a}_1,\hat{a}_2,\ldots,\hat{a}_K]$. The elements $a_k$ and $\hat{a}_k$ represent the true and predicted activity status of the $k$th user, respectively. 
Finally, each element of $\hat{\mat{a}}$ is compared with a threshold to determine the activity status of the users. The detection threshold is selected to ensure the desired false alarm rate.

\section{Numerical Results}\label{sec:results}
This section describes the simulation setup and provides the performance of the proposed approach.

\subsection{Simulation Setup}
\begin{table}[tb]
\centering
\caption{Simulation parameters}
\label{tab:sim_para}
\begin{tabular}{l| l}
\toprule
 Parameters & Value\\
 \midrule
Area side ($D$) & 1000 m \\
No. of \acp{AP} ($M$) & 20 \\
No. of users ($K$) & 200 \\
No. of antennas ($N$) & 1,2,3 \\
Pilot sequence length ($\tau$) & $40$\\
Carrier frequency ($f_{\mathrm{GHz}}$) & $1.9$ GHz \\
Shadowing intensity ($\sigma_\mathrm{sh}$) & $5.9$ \\
Transmit power ($\rho$) & $200$ mW \\
Noise power ($\sigma_{\mathrm{n}}^2$) & $-109$ dBm \\
User activation probability ($\epsilon$) & 0.1 \\
 \bottomrule
\end{tabular}
\end{table}
Throughout the simulations, we consider an industrial scenario with the following path-loss model  \cite{3GP:22}
\begin{align}\label{eq:pl}
    &\mathrm{PL}_{mk}[\mathrm{dB}] =  32.40 + 23\log_{10} d_{mk} + 20 \log_{10}f_{\mathrm{GHz}}  
\end{align}
where $f_{\mathrm{GHz}}$ is the carrier frequency in GHz and $d_{mk}$ is the distance between the $m$th \ac{AP} and $k$th user in meters. We consider an area of $1000 \times 1000$ m\textsuperscript{2} (i.e., $D=1000\,$m), and a simulation setting with $M=20$ \acp{AP} and $K=200$ users, utilizing a pilot sequence of length $\tau = 40$ symbols. Each device activates with probability $\epsilon = 0.1$. The complete set of simulation parameters can be found in Table~\ref{tab:sim_para}.

We employ $3 \times 10^6$ samples for the training dataset, and $10^4$ samples each for the validation and test datasets. We initialize the network with Glorot initialization \cite{GloBen:10}. The network is trained using mini-batches of $B = 256$ samples for $10$ epochs adopting the ADAM optimizer \cite{KinBa:15}. The simulator's code is available on GitHub \cite{aud-in-cf-mmimo:S24}.

\subsection{Performance Evaluation}
For performance evaluation, we consider the probability of detection (recall) and the false alarm rate. Probability of detection is defined as $ \textrm{R} =  \mathrm{TP}/{(\mathrm{TP + FN})}$ whereas false alarm rate is given by $\textrm{FA} = \mathrm{FP}/{(\mathrm{FP + TN})}$. Here, $\mathrm{TP}$, $\mathrm{FP}$, $\mathrm{TN}$, and $\mathrm{FN}$ denote the true positive, false positive, true negative, and false negative, respectively. 

\begin{figure}[tb]
   \centering
    \input{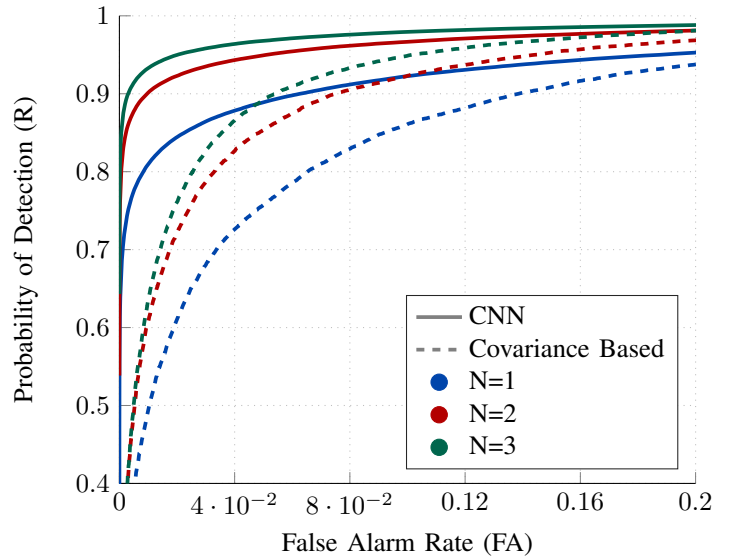}
    \caption{Receiver operating characteristics of CNN and Covariance-based approaches for different numbers of antennas $N$ with 20 \acp{AP} and users $K=200$.}
    \label{fig:plot1}
\end{figure}
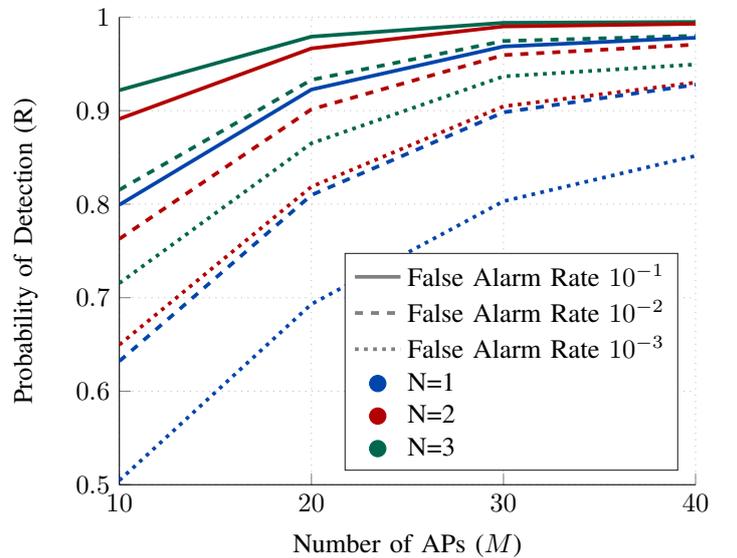
\begin{figure}[tb]
   \centering
%
%
\definecolor{mycolor1}{rgb}{0.00000,0.27000,0.67000}%
\definecolor{mycolor2}{rgb}{0.00000,0.40000,0.30000}%
\begin{tikzpicture}

\begin{axis}[%
width=0.856\hsize,
height=0.7\hsize,
at={(0\hsize,0\hsize)},
scale only axis,
xmin=10,
xmax=40,
xtick={ 0, 10, 20, 30, 40},
xlabel={Number of APs ($M$)},
ymin=0.5,
ymax=1,
ytick={0.5, 0.6, 0.7, 0.8, 0.9,   1},
ylabel={Probability of Detection ($\textrm{R}$)},
axis background/.style={fill=white},
axis x line*=bottom,
axis y line*=left,
xmajorgrids,
ymajorgrids,
grid style={dotted},
legend style={at={(0.97,0.03)}, anchor=south east, legend cell align=left, align=left, draw=white!15!black}
]
\addplot [color=gray, line width=1.4pt]
  table[row sep=crcr]{%
10	0.5\\
};
\addlegendentry{False Alarm Rate $10^{-1}$}

\addplot [color=gray, dashed, line width=1.4pt]
  table[row sep=crcr]{%
10	0.5\\
};
\addlegendentry{False Alarm Rate $10^{-2}$}

\addplot [color=gray, dotted, line width=1.4pt]
  table[row sep=crcr]{%
10	0.5\\
};
\addlegendentry{False Alarm Rate $10^{-3}$}

\addplot[only marks, mark=*, mark options={}, mark size=3.0619pt, draw=mycolor1, fill=mycolor1] table[row sep=crcr]{%
x	y\\
-1	-1\\
};
\addlegendentry{N=1}

\addplot[only marks, mark=*, mark options={}, mark size=3.0619pt, draw=black!30!red, fill=black!30!red] table[row sep=crcr]{%
x	y\\
-1	-1\\
};
\addlegendentry{N=2}

\addplot[only marks, mark=*, mark options={}, mark size=3.0619pt, draw=mycolor2, fill=mycolor2] table[row sep=crcr]{%
x	y\\
-1	-1\\
};
\addlegendentry{N=3}

\addplot [color=mycolor2, dotted, line width=1.4pt, forget plot]
  table[row sep=crcr]{%
10	0.715498439250844\\
20	0.865080240169199\\
30	0.936630009226704\\
40	0.94939237806252\\
};
\addplot [color=mycolor2, dashed, line width=1.4pt, forget plot]
  table[row sep=crcr]{%
10	0.815291339843128\\
20	0.932916010945945\\
30	0.974469464601881\\
40	0.979991287497562\\
};
\addplot [color=mycolor2, line width=1.4pt, forget plot]
  table[row sep=crcr]{%
10	0.921832479590201\\
20	0.979150586890931\\
30	0.993910376299844\\
40	0.994822570388507\\
};
\addplot [color=black!30!red, dotted, line width=1.4pt, forget plot]
  table[row sep=crcr]{%
10	0.649716864094763\\
20	0.818645189098163\\
30	0.904635794618585\\
40	0.929984527797387\\
};
\addplot [color=black!30!red, dashed, line width=1.4pt, forget plot]
  table[row sep=crcr]{%
10	0.762866175764366\\
20	0.901416356765537\\
30	0.959402508665619\\
40	0.97059280066896\\
};
\addplot [color=black!30!red, line width=1.4pt, forget plot]
  table[row sep=crcr]{%
10	0.891192772530815\\
20	0.966555737096918\\
30	0.990045136031519\\
40	0.992849739376005\\
};
\addplot [color=mycolor1, dotted, line width=1.4pt, forget plot]
  table[row sep=crcr]{%
10	0.504652233071873\\
20	0.693087617653838\\
30	0.803102167028257\\
40	0.851617070496161\\
};
\addplot [color=mycolor1, dashed, line width=1.4pt, forget plot]
  table[row sep=crcr]{%
10	0.632113414438933\\
20	0.809729057867145\\
30	0.898207027256177\\
40	0.927731294407472\\
};
\addplot [color=mycolor1, line width=1.4pt, forget plot]
  table[row sep=crcr]{%
10	0.799313670561872\\
20	0.922561470685523\\
30	0.96858931198723\\
40	0.977963377446635\\
};
\end{axis}
\end{tikzpicture}%
    \caption{Detection probability with fixed false alarm rates for different number of antennas and \acp{AP}.}
    \label{fig:plot2}
\end{figure}
As a performance benchmark, we compare our \ac{CNN} with the algorithm presented in \cite{HagJunCai:18}, which does not require the estimation of the large-scale fading coefficients. While in \cite{HagJunCai:18} the algorithm operates in a single-cell scenario, we extend it to a cell-free environment by introducing an additional fusion step to merge user activity detection results from individual \acp{AP}. Final detection is thus obtained by performing the union of the decisions taken by the \acp{AP} individually.

The \ac{ROC} curves for the \ac{DL} and covariance-based approaches \cite{HagJunCai:18} are reported in Fig.~\ref{fig:plot1}. It is evident from Fig.~\ref{fig:plot1} that the performance improves as the number of antennas increases at each \ac{AP} for both approaches. For instance, we obtain false alarm rates equal to $0.06$, $0.01$, and $0.0036$ for a fixed detection probability of $0.90$ from the \ac{CNN}, when the number of antennas is $1$, $2$, and $3$, respectively. For $N=1$, a significant performance gap between the proposed approach and the covariance-based one can be observed. Specifically, at $\textrm{FA}=0.04$, the covariance-based approach achieves a detection probability of $0.73$, whereas the \ac{CNN} achieves $0.88$.

To assess the impact of distributing a large number of antennas within the network area on the proposed algorithm, we evaluate its performance varying the number of network \acp{AP}. In Fig.~\ref{fig:plot2}, we depict the probability of detection of active users for a fixed false alarm rate of $10^{-1}$, $10^{-2}$, and $10^{-3}$. The results demonstrate a clear improvement in performance when more \acp{AP} are deployed in the area. For instance, increasing the number of \acp{AP} from $10$ to $40$ with $N=1$ and $\textrm{FA}=10^{-3}$ leads to a $70\%$  jump in detection probability. Likewise, with the false alarm rate of $10^{-1}$ and $M=10$, we can observe an increase in detection rate from $0.79$ to $0.92$ by varying the number of antennas from $1$ to $3$. Lower false alarm rates inherently result in lower detection probabilities. Precisely, for $M=20$ and $N=3$, we obtain the detection probabilities $0.98$, $0.93$, and $0.86$ for false alarm rates of $10^{-1}$, $10^{-2}$, and $10^{-3}$, respectively.

\section{Conclusion}\label{sec:conclusion}
In this paper, we propose a \ac{DL}-based method for user activity detection in a \ac{CF-mMIMO} scenario. 
The proposed method is based on a \ac{CNN} that employs channel estimates to determine the activity status of each user. The algorithm is blind, allowing user activity detection without the need to estimate the large-scale fading coefficients between the users and the \acp{AP}. The performance of the \ac{CNN} is compared to that of a covariance-based method properly modified to operate in a cell-free scenario.  
Our \ac{CNN} exhibits remarkable detection probability, surpassing the covariance-based benchmark. The simulation results underscore the advantages of cell-free network architectures for user activity detection.

\section*{Acknowledgements}
Supported by the European Union under the Italian National Recovery and Resilience Plan of NextGenerationEU, partner- ship on “Telecommunications of the Future” (PE00000001 - “RESTART”).

\bibliographystyle{IEEEtran}
\bibliography{Files/IEEEabrv,Files/Refs,Files/ETbib}

\end{document}